\documentclass[twocolumn,showpacs,prl]{revtex4}

\usepackage{amsmath,amssymb}
\usepackage{graphicx}

\draft

\begin{document}

\title{Manipulating Quantum States of Molecules Created via Photoassociation of Bose-Einstein Condensates}

\author{Xiao-Ting Zhou$^{a}$, Xiong-Jun Liu$^{b}$\footnote{Electronic address: phylx@nus.edu.sg},
Hui Jing$^{c,d}$, C. H. Lai$^{b}$ and C. H.
Oh$^{b}$\footnote{Electronic address: phyohch@nus.edu.sg}}
\affiliation{a. Theoretical Physics Division, S. S. Chern
Institute of
Mathematics, Nankai University, Tianjin 300071, P.R.China\\
b. Department of Physics, National University of Singapore, 2
Science Drive 3, Singapore 117542\\
c. Department of Physics, The University of Arizona, 1118 East 4th
Street, Tucson, AZ 85721\\
d. State Key Lab of Magnetic Resonance and Atomic and Molecular
Physics, Wuhan Institute of Physics and Mathematics, CAS, Wuhan
430071, P.R. China}

\date{\today}

\begin{abstract}
We show the quantum state transfer technique in two-color
photoassociation (PA) of a Bose-Einstein condensate, where a
quantized field is used to couple the free-bound transition from
atom state to excited molecular state. Under the weak excitation
condition, we find that quantum states of the quantized field can
be transferred to the created molecular condensate. The
feasibility of this technique is confirmed by considering the
atomic and molecular decays discovered in the current PA
experiments. The present results allow us to manipulate quantum
states of molecules in the photoassociation of a Bose-Einstein
condensate.
\end{abstract}

\maketitle

Over the last few years, a particularly important development in
ultracold quantum systems has been the ability to form and
manipulate quantum degenerate molecular gases via photoassociation
(PA) in Bose-Einstein Condensates (BECs). Besides the creation of
molecular quantum gases with Feshbach resonances \cite{Feshbach},
a more general method, say, a stimulated optical Raman transition
(STIRAP) has been employed to directly produce deeply bound
molecules in recent years \cite{optical1,optical2}. STIRAP was
proposed as a promising way for a fast, efficient and robust
process to convert a BEC of atoms into a molecular condensate
\cite{molecular1,molecular2,molecular3}. The central idea for this
kind of STIRAP is the realization of the dark superposition state
of a BEC of atoms and a BEC of molecules, which has been observed
in recent experiments \cite{experiment}.

Although PA process has been widely studied in both theoretical
and experimental aspects, the possibility of manipulating quantum
states of the created cold molecular gas is not well discovered in
such process. Control of quantum states of molecular gas may be
achieved by realizing quantum state transfer from photons to
molecules formed in the PA process. Most recently, one of us
studied the possibility of quantum conversion between light and
molecules via coherent two-color PA by using a single-mode
associating light \cite{Hui}. Here we shall study further the
validity of quantum state transfer from a quantized associating
light to molecular gas in the PA process.

In this letter, we identify the quantum state transfer process in
two-color PA of a Bose-Einstein condensate by quantizing
associating light. Under the weak excitation condition, we show
quantum states of the quantized field can be fully transferred to
the created molecular condensate. The multi-mode associating light
is applied and the effect of atomic/molecular decays is discussed
in detail in this model. The present result confirms the
possibility of quantum state transfer in the current PA
experiments, and will have wide applications to, e.g. quantum
information science with cold molecules \cite{computing}.

\begin{figure}[ht]
\includegraphics[width=0.7\columnwidth]{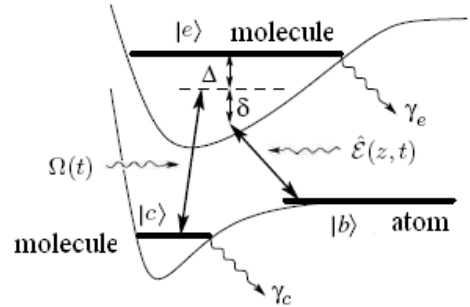}
\caption{Level scheme of two-color PA process induced via a
quantized and a classical field. $\delta$ and $\Delta$ denote the
detunings.} \label{fig1}\end{figure} Turning to the situation of
Fig. 1, we consider the quasi one-dimensional model of N identical
atoms that have condensed into the same one-particle state with
wave vector $\bold k=0$. Through a quasi-one dimensional quantized
associating field $\hat{E}(z, t)=\sqrt{\frac{\hbar\nu}{2\epsilon_0
L}}\hat{\cal E}(z, t)\exp(i\frac{\nu}{c}(z-ct))$, PA then removes
two atoms from the state $|b\rangle$, creating a molecule in the
excited state $|e\rangle$. Here $\hat{\cal E}(z, t)$ is slowly
varying amplitude, $L$ is the quantization length in the $z$
direction and $\nu$ the frequency of associating field. Including
a second coherent classical laser characterized by the coupling
strength (Rabi frequency) $\Omega(t)$, bound-bound transitions
remove excited molecules from state $|e\rangle$ and create stable
molecules in state $|c\rangle$. In second-quantized notation, we
denote the boson field operators for atoms, primarily
photoassociated molecules, and stable molecules, respectively, by
$\hat\phi_b, \hat\phi_e$ and $\hat\phi_c$. Atomic and quantized
field operators satisfy the following commutation relations:
$[\hat{\phi_{\mu}}(z, t), \hat{\phi_{\nu}}^{\dagger}(z',
t)]=\delta_{\mu\nu} \delta (z-z'), \ [\hat{E} (z, t),\ \ \hat{E}
^\dagger (z',t)] =(\frac{\hbar\nu}{\epsilon_0})\delta (z-z')$.
With these definitions, the interaction Hamiltonian of our system
can be given by
\begin{eqnarray}\label{eqn:Hamiltonian1a}
\hat{H}&=&-\int dz\bigr(\hbar g \hat{\cal E}(z,
t)\hat{\phi_{e}}^{\dagger} (z, t)\hat{\phi_{b}}(z,
t)\hat{\phi_{b}}(z, t)\nonumber\\
&+&\hbar \Omega \hat{\phi_{e}}^{\dagger} (z, t) \hat{\phi_{c}}(z,
t)+h.c.\bigr)+\hbar \Delta\int dz\hat{\phi_{e}}^{\dagger} (z, t)
\hat{\phi_{e}}(z, t)\nonumber\\
&+& \hbar \delta\int dz\hat{\phi_{b}}^{\dagger} (z, t)
\hat{\phi_{b}}(z, t),
\end{eqnarray}
where $g$ is the coupling coefficient between the atoms and
quantized field, $\Delta$ and $\delta$ represent the one- and
two-photon detunings, respectively. We note here $g$ has the units
of s$^{-1}$m$^{1/2}$, since here we consider quasi-one dimensional
system \cite{molecular3,experiment}.

The evolution of the quantum field can be described in the slowly
varying amplitude approximation by the propagation equation:
\begin{eqnarray}\label{eqn:light1}
(\frac{\partial}{\partial t}+c\frac{\partial}{\partial
z})\hat{\cal E}(z, t)= i gL\hat{\phi_{b}}^{\dagger} (z,
t)\hat{\phi_{b}}^{\dagger} (z, t)\hat{\phi_{e}} (z, t).
\end{eqnarray}
On the other hand, the evolution of atomic field operators is
governed by a set of Heisenberg equations
\begin{eqnarray}\label{eqn:field1}
\dot{\hat{\phi_{b}}} &=& -i\delta\hat\phi_b- \gamma_{b}
\hat{\phi_{b}}+i g \hat{\mathcal {E}}^{\dagger}
\hat{\phi_{b}^{\dagger}} \hat{\phi_{e}}, \nonumber\\
\dot{\hat{\phi_{e}}} &=& i \Delta \hat{\phi_{e}}-\gamma_{e}
\hat{\phi_{e}}++ i g \hat{\cal E} \hat{\phi_{b}} \hat{\phi_{b}}+i
\Omega \hat{\phi_{c}},\\
\dot{\hat{\phi_{c}}} &=& - \gamma_{c} \hat{\phi_{c}}+i \Omega
\hat{\phi_{e}},\nonumber
\end{eqnarray}
where $\gamma_{b}$,\ $\gamma_{c}$ and $\gamma_{e}$ denote the
decay rates of corresponding atomic or molecular states. With
these formulas we find the atom-molecule evolutions can be
described by
\begin{eqnarray}\label{eqn:field2}
\frac{\partial}{\partial t}({\hat{\phi_{b}}}^{\dagger {2}}
\hat{\phi_{e}}) &=& -(i\Delta+i2\delta+2 \gamma_{b}+\gamma_{e})
{\hat{\phi_{b}}}^{\dagger {2}} \hat{\phi_{e}} +i g \hat{\cal E}
{\hat{\phi_{b}}}^{\dagger {2}}{\hat{\phi_{b}}}^{2}  \nonumber\\
&& - i g {\hat{\cal E}^{\dagger}} {\hat{\phi_{b}}}^{\dagger}
\hat{\phi_{b}}{\hat{\phi_{e}}}^{\dagger} \hat{\phi_{e}} + i \Omega
{\hat{\phi_{b}}}^{\dagger {2}}\hat{\phi_{c}},\\
\frac{\partial}{\partial t}({\hat{\phi_{b}}}^{\dagger {2}}
\hat{\phi_{c}} ) &=& -(i2\delta+2\gamma_{b}+\gamma_{c})
{\hat{\phi_{b}}}^{\dagger {2}} \hat{\phi_{c}} - i g {\hat{\cal
E}^{\dagger}} {\hat{\phi_{b}}}^{\dagger}
\hat{\phi_{b}}{\hat{\phi_{e}}}^{\dagger}
\hat{\phi_{c}}\nonumber\\
&& +\ i \Omega{\hat{\phi_{b}}}^{\dagger {2}}\hat{\phi_{e}}.
\end{eqnarray}
It should be noted that we have disregarded the motion of the
trapped atoms and molecules. This is valid since the time scale
for coherent STIRAP is shorter than the time scale for the motion
of the atoms and/or molecules in the trap
\cite{molecular1,molecular2,molecular3}. Also, as long as laser
intensities permit STIRAP during a time interval much shorter than
the time scales for collisions between atoms and molecules,
collisions are negligible as well. The above two Eqs. can be
recast into
\begin{eqnarray}\label{eqn:field4}
{\hat{\phi_{b}}}^{\dagger {2}} \hat{\phi_{e}} &=&
-\frac{i}{\Omega}\frac{\partial}{\partial
t}({\hat{\phi_{b}}}^{\dagger {2}} \hat{\phi_{c}})
-\frac{i}{\Omega} (\gamma_{bc}+i2\delta){\hat{\phi_{b}}}^{\dagger
{2}}
\hat{\phi_{c}} \nonumber\\
&& + \frac{g}{\Omega} \hat{\cal
E}^{\dagger}{\hat{\phi_{b}}}^{\dagger}
\hat{\phi_{b}}{\hat{\phi_{e}}}^{\dagger} \hat{\phi_{c}}.
\end{eqnarray}
\begin{eqnarray}\label{eqn:field5}
{\hat{\phi_{b}}}^{\dagger {2}} \hat{\phi_{c}} &=&
-\frac{i}{\Omega}\frac{\partial}{\partial
t}({\hat{\phi_{b}}}^{\dagger {2}} \hat{\phi_{e}})
-\frac{i}{\Omega}
(\gamma_{be}+i\Delta+i2\delta){\hat{\phi_{b}}}^{\dagger
{2}}\hat{\phi_{e}} \nonumber\\
&&+\frac{g}{\Omega}\hat{\cal
E}^{\dagger}{\hat{\phi_{b}}}^{\dagger}
\hat{\phi_{b}}{\hat{\phi_{e}}}^{\dagger} \hat{\phi_{c}}
-\frac{g}{\Omega} \hat{\cal E}{\hat{\phi_{b}}}^{\dagger
{2}}{\hat{\phi_{b}}}^{2}.
\end{eqnarray}
Here $\gamma_{bc}=2\gamma_{b} + \gamma_{c}$ and
$\gamma_{be}=2\gamma_{b}+\gamma_{e}$ are transversal decay rates.
From Eqs. (\ref{eqn:field4}) and (\ref{eqn:field5}), we further
obtain
\begin{eqnarray}\label{eqn:weak1}
{\hat{\phi_{b}}}^{\dagger {2}} \hat{\phi_{c}}
&=&\frac{\Omega^2}{\Omega^2+(\gamma_{be}+i2\delta+i\Delta)(\gamma_{bc}+i2\delta)}\biggl(-\frac{g\hat{\cal
E}}{\Omega} {\hat{\phi_{b}}}^{\dagger
{2}}{\hat{\phi_{b}}}^{2}\nonumber\\
&-&\frac{i}{\Omega} \frac{\partial}{\partial
t}\bigl[-\frac{1}{\Omega}(1+\gamma_{be}+\gamma_{bc}+4i\delta+i\Delta){\hat{\phi_{b}}}^{\dagger
{2}} \hat{\phi_{c}}\nonumber\\
&+&\frac{1}{\Omega} (\Delta-i
\gamma_{be}){\hat{\phi_{b}}}^{\dagger
{2}}\hat{\phi_{e}}+\frac{g\hat{\cal
E}^{\dagger}}{\Omega}{\hat{\phi_{b}}}^{\dagger}
\hat{\phi_{b}}{\hat{\phi_{e}}}^{\dagger}
\hat{\phi_{c}}\bigl]\nonumber \\
&+&\frac{g\hat{\cal
E}^{\dagger}}{\Omega}{\hat{\phi_{b}}}^{\dagger}
\hat{\phi_{b}}{\hat{\phi_{e}}}^{\dagger} \hat{\phi_{c}}\biggl).
\end{eqnarray}

The above equation is hard to tackle. However, according to the
present experiments \cite{experiment}, we can simplify it with
several approximations. Firstly, we shall consider the
weak-excitation condition, i.e. only a small ratio of atoms are
converted into molecules, thus the last term in the right-hand
side (r.h.s.) of above formula is very small and can be safely
neglected. Secondly, we assume $\Omega$ changes sufficiently
slowly with time so that adiabatic condition is fulfilled. For
this only the first term in the r.h.s. of (\ref{eqn:weak1}) plays
important role in the dynamical evolution of atomic fields. On the
other hand, in this paper we shall first neglect the effect of
two-photon detuning $(\delta)$ and atomic-molecular decay
$(\gamma_{bc})$ that will be discussed later in detail. By
introducing a normalized time $\tau=t/T$, where $T$ is a
characteristic time scale, and expanding the r.h.s. of
(\ref{eqn:weak1}) in powers of $1/T$, we find in lowest
nonvanishing order,
\begin{eqnarray}\label{eqn:weak2}
{\hat{\phi_{b}}}^{\dag {2}} \hat{\phi_{c}} &\approx& - \frac{g
\hat{\cal E}}{\Omega} {\hat\phi_{b}}^\dagger\
{\hat\phi_{b}}^{\dagger}\ \hat{\phi_{b}}\ \hat\phi_{b}.
\end{eqnarray}
Thus Eq. (\ref{eqn:field4}) can be recast into
\begin{eqnarray}\label{eqn:weak3}
{\hat{\phi_{b}}}^{\dag {2}} \hat{\phi_{e}}\approx i\frac{g
n^{2}}{\Omega^{2}} \frac{\partial}{\partial t}\hat{\cal E}(z, t)
-i \frac{g n^{2}}{\Omega^{3}} \frac{\partial\Omega(t)}{\partial
t}\hat{\cal E}(z, t).
\end{eqnarray}
Here $n^2={\hat{\phi_{b}}}^{\dagger {2}}{\hat{\phi_{b}}}^{2}$ is
the atomic BEC density that is assumed to be constant in weak
excitation case. Substituting this result into Eq.
(\ref{eqn:light1}) yields
\begin{eqnarray}\label{eqn:light2}
(\frac{\partial}{\partial t} &+& \frac{c}{1+\frac{g^{2}n
N}{\Omega^{2}}}\frac{\partial}{\partial
z})\hat{\cal E}(z,t)= \nonumber\\
&&= \frac{1}{1+\frac{g^{2}n N}{\Omega^{2}}}\frac{g^{2}n
N}{\Omega^{3}} \frac{\partial\Omega}{\partial t} \hat{\cal E}(z,
t),
\end{eqnarray}
where $N=\langle{\hat{\phi_{b}}}^{\dagger}{\hat{\phi_{b}}}\rangle$
is the total atomic number in the interacting region. Noting that
$\Omega(t)$ is time dependent, the r.h.s. of above equation
(\ref{eqn:light2}) represents an adiabatic Raman absorption or
enhancement of the quantized field. The group velocity of the
associating field is given by $v_{g} = c \cos^{2}\theta =
c/\bigl(1+\frac{\tilde{g}^{2}N^{2}}{\Omega^{2}}\bigl)$. Here
$\tilde{g}=gn^{1/2}N^{-1/2}$ is related to atomic density
$n^{1/2}$ \cite{experiment} and can be regarded as constant in the
weak-excitation condition. The mixing angle is defined according
to $\tan^2\theta=\tilde{g}^2N^2/\Omega^2(t)$ that is governed by
the atomic number and the strength of external classical field.
Different from the usual electromagnetically induced transparency
(EIT) case where the group velocity of associating light is
proportional to $N^{-1}$ \cite{EIT}, here the group velocity of
quantized field is proportional to $N^{-2}$ due to the
nonlinearity in the Hamiltonian (\ref{eqn:Hamiltonian1a}). When
the classical field is adiabatically turned off so that the mixing
angle $\theta\rightarrow\pi/2$, one has $v_g\rightarrow0$, which
means the initial photon wave packet is decelerated to a full
stop. Equation (\ref{eqn:light2}) has the following simple
solution
\begin{eqnarray}\label{eqn:light3}
\hat{\mathcal {E}}(z,t)=\frac{\cos\theta(t)}{\cos\theta(0)}\
\hat{\mathcal {E}}(z-\int_0^tv_gdt',0).
\end{eqnarray}
On the other hand, from Eq. (\ref{eqn:weak2}), one can easily find
\begin{eqnarray}\label{eqn:state3}
\hat{\phi_{c}}(z,t)\approx -
\frac{gn}{\Omega(0)}\sqrt{\frac{\Omega^2(0)+\tilde{g}^2N^2}{\Omega^2(t)+\tilde{g}^2N^2}}\hat{\mathcal
{E}}(z-\int_0^tv_gdt',0).
\end {eqnarray}

With the expression (\ref{eqn:state3}) one can readily reach a
full quantum state transfer from photons to created molecules via
PA process. For example, if initially a strong classical field is
applied so that $\Omega^2\gg \tilde{g}^2N^2$ or $\theta(0)=0$, one
has $\hat\phi_c(z,0)=0$. Then, after $\Omega$ is adiabatic turned
off, i.e. $\theta(t)=\pi/2$, from (\ref{eqn:state3}) we have
$L\hat{\phi_{c}}(z,t)= -\hat{\mathcal {E}}(z-\int_0^tv_gdt',0)$,
say, quantum states of the associating light are fully transferred
into the created molecular BEC. The similar result is well-known
in quantum memory technique with EIT. Nevertheless, identification
of quantum state transfer in the present model suggests a novel
technique to create molecules in non-classical states via a PA
process.

In above discussions, we have ignored the decay
$\gamma_{bc}=2\gamma_b+\gamma_c$ of ground atomic and molecular
states. However, recent experiments \cite{experiment} indicate
$\gamma_c$ is not very small in a PA process. As a result, it is
important to further verify the validity of quantum state transfer
technique when decay of atomic and molecular states is included.
Noting that one- and two-photon detunings can be adjusted in
experiment \cite{experiment}, we would like to consider here small
detuning case so that $\delta^2\ll\gamma_{bc}^2$ and $\delta^2,
\Delta^2\ll\gamma_{be}^2$. The lowest order in Eq.
(\ref{eqn:weak1}) then reads
\begin{eqnarray}\label{eqn:decay1}
{\hat{\phi_{b}}}^{\dagger {2}} \hat{\phi_{c}} &=&\frac{-g
{\hat{\phi_{b}}}^{\dagger
{2}}{\hat{\phi_{b}}}^{2}}{\Omega+\frac{1}{\Omega}\gamma_{be}\gamma_{bc}}\hat{\cal
E}(z,t).
\end{eqnarray}
Substituting this result into Eq. (\ref{eqn:field4}) and then into
Eq. (\ref{eqn:light1}) yields
\begin{widetext}
\begin{eqnarray}\label{eqn:light4}
\biggl(\bigl(1+
\frac{\tilde{g}^2N^2}{\Omega^2+\gamma_{be}\gamma_{bc}}\bigl)\frac{\partial}{\partial
t}+c\frac{\partial}{\partial z}\biggl)\hat{\cal
E}(z,t)=-\frac{\tilde{g}^{2}N^2}{\Omega} \frac{\partial}{\partial
t}\biggl(\frac{1}{\Omega+\frac{1}{\Omega}\gamma_{be}\gamma_{bc}}\biggl)\hat{\cal
E}(z,
t)-\frac{\tilde{g}^2N^2\gamma_{bc}}{\Omega^2+\gamma_{bc}\gamma_{be}}\hat{\cal
E}(z, t).
\end{eqnarray}
\end{widetext}

Thus the quantum field propagates with the group velocity
\begin{eqnarray}\label{eqn:velocity1}
v_g=c\bigl(1+\frac{\tilde{g}^2N^2}{\Omega^2+\gamma_{be}\gamma_{bc}}\bigl)^{-1},
\end{eqnarray}
which approaches
$\gamma_{be}\gamma_{bc}c/(\tilde{g}^2N^2+\gamma_{be}\gamma_{bc})$
when $\Omega\rightarrow0$. The presence of atomic decay results in
a nonzero group velocity even when the classical field is turned
off. For weak associating field case, typical parameter values can
be taken as \cite{experiment}
$\gamma_{be}\approx2.0\times10^{7}$s$^{-1}$,
$\gamma_{bc}\approx5.0\times10^{3}$s$^{-1}$,
$N\approx3.0\times10^6$ and $\tilde{g}\approx50$ s$^{-1}$. One can
then evaluate the limit of group velocity by $v^c_g\approx1.33$
km$\cdot$s$^{-1}$. Furthermore, if we consider a larger number of
trapped atoms, say $N\sim10^8$ \cite{BEC1}, the limit of group
velocity will be decreased by four orders and yields
$v_g^c\approx0.13$ m$\cdot$s$^{-1}$. The numerical result
indicates the associating light can still reach an approximate
``stop" for the large atomic number case.

The first term in the r.h.s. of above equation (\ref{eqn:light4})
represents an adiabatic Raman absorption or enhancement of the
quantized field, while the second term represents a decay. The
solution can be given by
\begin{eqnarray}\label{eqn:solution1}
\hat{\mathcal {E}}(z,t)=\hat{\mathcal {E}}(z-\int_0^tv_gdt',0)\
e^{\int_0^tdt'\alpha(t')}e^{-\int_0^tdt'\Gamma(t')},
\end{eqnarray}
where
\begin{eqnarray}\label{eqn:}
\int_0^t\alpha(t')dt'&=&-\int_0^t\frac{\tilde{g}^2N^2}{\Omega}\frac{\partial_{t'}\Omega
(\Omega^2+\gamma_{be}\gamma_{bc})-2\Omega^2\partial_{t'}\Omega}{(\Omega^2+\gamma_{be}\gamma_{bc})^2}\times\nonumber\\
&&\times\frac{dt'}{1+\frac{\tilde{g}^2N^2}{\Omega^2+\gamma_{be}\gamma_{bc}}},\nonumber
\end{eqnarray}
$$\int_0^t\Gamma(t')dt'=\int_0^t\frac{\tilde{g}^2N^2\gamma_{bc}dt'}{\Omega^2+\gamma_{be}
\gamma_{bc}+\tilde{g}^2N^2}.$$

The above integral can be calculated straightforwardly. Similar to
the former discussion, we assume initially the classical field is
strong so that
$\Omega^2(0)\gg\tilde{g}^2N^2,\gamma_{be}\gamma_{bc}$, while
finally $\Omega^2(t)\ll\tilde{g}^2N^2$. We then find from Eq.
(\ref{eqn:solution1})
\begin{eqnarray}\label{eqn:solution2}
\hat{\mathcal {E}}(z,t)&=&\hat{\mathcal
{E}}(z-\int_0^tv_gdt',0)\biggl(\frac{\Omega(0)}{\Omega(t)}\biggl)^f\biggl(\frac{\gamma_{be}\gamma_{bc}+\Omega^2(t)}{\Omega^2(0)}\biggl)
\times\nonumber\\
&&\times\biggl(\frac{\Omega(0)}{\gamma_{be}\gamma_{bc}+\tilde{g}^2N^2}\biggl)^h\exp\bigl(-\int_0^t\Gamma(t')dt'\bigl),
\end{eqnarray}
where $f=\tilde{g}^2N^2/(\gamma_{be}\gamma_{bc}+\tilde{g}^2N^2)$
and
$h=1/2+\gamma_{be}\gamma_{bc}/(2\gamma_{be}\gamma_{bc}+2\tilde{g}^2N^2)$.
The above equation reduces into the formula (\ref{eqn:light3}) if
$\gamma_{be}\gamma_{bc}\rightarrow0$. Based on the previous
parameters used in calculating the limit of group velocity, one
can see $\tilde{g}^2N^2\gg\gamma_{be}\gamma_{bc}$. Thus we have
$f\approx1$ and $h\approx1/2$, and the created molecular BEC is
given by
\begin{eqnarray}\label{eqn:solution3}
L\hat{\phi_{c}}=-\hat{\mathcal
{E}}(z-\int_0^tv_gdt',0)\exp\bigl(-\int_0^t\Gamma(t')dt'\bigl).
\end{eqnarray}
This formula shows quantum state transfer technique from quantized
associating field to created molecules is still valid when
$|1-\exp\bigl(-\int_0^t\Gamma(t')dt'\bigl)|\leq|1-\exp\bigl
(-\int_0^t\gamma_{bc}dt'\bigl)|\ll1$. Therefore, to obtain a
sufficient quantum state transfer, the time duration of a PA
process should satisfy $t\ll t_{max}\sim\gamma_{bc}^{-1}$. For
$^{87}Rb$ atomic-molecular BECs, the decay $\gamma_{bc}$ is about
$10^3$s$^{-1}$ for the weak associating case. Then the time limit
is about $t_{max}\sim1.0$ ms, which is in the same order with
storage time of EIT quantum memory \cite{EIT experiment}. Such
requirement can also be expressed that the propagation depth of
associating light satisfy $z\ll
z_{max}=\int_0^{t_{max}}v_gdt\sim\gamma_{be}c/\tilde{g}^2N^2$
during the PA process. With typical values of parameters one can
find $z_{max}$ is about $1.0\sim 10$ mm. For BECs, the spatial
scale of the interaction region is about $0.1\sim0.01$ mm
\cite{BEC}. Thus the requirement can be generally satisfied if
only the classical field is turned down to be
$\Omega^2(t)\ll\tilde{g}^2N^2$ before the quantized field
propagates through the atomic-molecular BEC in PA experiments.
These results show that even for the case of nonzero atomic
decays, we can still reach a novel technique to manipulate quantum
states of created molecules via a quantized associating field as
long as the time of PA process is small enough. Now that the
mapping technique can be applied to separate Raman transitions at
the same time, it is also possible to transfer entanglement from a
pair of associating light beams as generated, e.g. in parametric
down-conversion to a pair of created molecular BECs.

As a final remark, we note that the above analysis involves the
weak-excitation approximation, valid when ratio of molecular
number and atomic number is very small. Making use of Eq.
(\ref{eqn:solution3}) one finds
$\langle\hat{\phi_c}^\dag\hat{\phi_c}\rangle\approx\langle\hat{\cal
E}^\dag\hat{\cal E}\rangle$. If the initial number density of
associating photons is much less than the number density of atoms,
the formed molecular number is always much smaller than the atomic
number, and the weak-excitation condition is valid during the PA
process. It is also noteworthy that such condition is true in the
present PA experiments \cite{experiment}.

In conclusion, We have shown the quantum state transfer process in
two-color PA of a Bose-Einstein condensate, where a quantized
field is used to couple the free-bound transition from atom state
to excited molecular state. Under the weak excitation condition,
we find quantum states of the quantized field can be transferred
to the created molecular condensates. Effects of atomic and
molecular decays in the present model are discussed in detail. Our
results show that quantum state transfer proposed here is
achievable in the current PA experiments, and will have wide range
of applications to quantum information science. For example, if
the created molecular state is untrapped, we may create molecular
lasers with controllable quantum states. Besides, if considering a
squeezed associating light in the PA process, one may study the
molecule-light entanglement \cite{entanglement}, and so on.

This work is supported by NUS academic research Grant No. WBS:
R-144-000-189-305, and by NSF of China under grants No. 10575053.






\begin{thebibliography}{99}
\bibitem{Feshbach} D. Kleppner, Phys. Today 57, No. 8, 12 (2004).

\bibitem{optical1} R. Wynar, R. S. Freeland, D. J. Han, C. Ryu, and D. J.
Heinzen, Science 287, 1016 (2000).

\bibitem{optical2} B. Laburthe Tolra, C. Drag, and P. Pillet, Phys. Rev. A 64,
061401 (2001).

\bibitem{molecular1} A. Vardi, D. Abrashkevich, E. Frishman, and M. Shapiro,
J. Chem. Phys. 107, 6166 (1997); A. Vardi et al., Phys. Rev. A 64,
063611 (2001).

\bibitem{molecular2} P. S. Julienne, K. Burnett, Y. B. Band, and W. C. Stwalley,
Phys. Rev. A 58, R797 (1998); J. Javanainen and M. Mackie, Phys.
Rev. A 58, R789 (1998); J. J. Hope et al., Phys. Rev. A 63, 043603
(2001).

\bibitem{molecular3} P. D. Drummond, K.V. Kheruntsyan, D. J. Heinzen, and R. H. Wynar,
Phys. Rev. A 65, 063619 (2002); 71, 017602 (2005); B. Damski et
al., Phys. Rev. Lett. 90, 110401 (2003); H.Y. Ling, H. Pu, and B.
Seaman, Phys. Rev. Lett. 93, 250403 (2004).

\bibitem{experiment} K. Winkler, G. Thalhammer, M. Theis, H. Ritsch, R. Grimm, and J. Hecker
Denschlag, Phys. Rev. Lett. 95, 063202 (2005); R. Dumke, J. D.
Weinstein, M. Johanning, K. M. Jones, and P. D. Lett,
cond-mat/0508077.

\bibitem{Hui} H. Jing and M. -S. Zhan, to appear in Euro. Phys. J.
D, quant-ph/0512149.

\bibitem{computing} M. Baranov,  L. Dobrek, K. G\'{o}ral, L. Santos, and
M. Lewenstein, Phys. Scr. T102, 74 (2002); D. DeMille, Phys. Rev.
Lett. 88, 067901 (2002).

\bibitem{EIT} S. E. Harris et al., Phys. Rev. A 46, R29 (1992);
M. O. Scully and M. S. Zubairy, Quantum Optics (Cambridge
University Press, Cambridge 1999); M. Fleischhauer and M. D.
Lukin, Phys. Rev. Lett. 84, 5094 (2000); Phys. Rev. A 65, 022314
(2002); X. J. Liu, H. Jing, X. Liu and M. L. Ge, Phys. Lett. A
355, 437 (2006).

\bibitem{BEC1} See e.g. B. Desruelle, V. Boyer, S. G. Murdoch, G. Delannoy, P. Bouyer, and A.
Aspect, Phys. Rev. A 60, R1759 (1999).

\bibitem{EIT experiment} D. F. Phillips et al., Phys. Rev. Lett. 86, 783
(2001); C. Liu et al., Nature (London) 409, 490 (2001).

\bibitem{BEC} F. Dalfovo, S. Giorgini, L. P. Pitaevskii
and S. Stringari, Rev. Mod. Phys. 71, 463 (1999), and references
therein.

\bibitem{entanglement} S. A. Haine, M. K. Olsen and J. J. Hope, Phys. Rev. Lett. 96, 133601
(2006).
\end{thebibliography}
\end{document}